# Transformation optics with Fabry-Pérot resonances

*M. M. Sadeghi [a], Sucheng Li [a], Lin Xu [a], Bo Hou, and Huanyang Chen*

*School of Physical Science and Technology, Soochow University, Suzhou 215006, China*

**Transformation optics [1,2] is a powerful tool to design various novel devices [3], such as invisibility cloak [4]. Fantastic effects from this technique are usually accompanied with singular mappings, resulting in challenging implementations and narrow bands of working frequencies. Here in this article, we find that Fabry-Pérot resonances can be used to design various transformation optical devices that are not only easy to realize but also can work well for a set of resonant frequencies (multiple frequencies). As an example, we fabricate a prototype for a cylindrical concentrator for microwaves.**

Transformation optics [1,2] has become a juicy topic since it emerged in 2006, as a fantastic tool to design various devices with novel functionalities. Readers can find the progress in a recent review [3]. One of the key concerns however in this field is that, the devices designed for the intriguing effect usually come from some singular mappings. The singular material parameters are thereby required, regardless the difficulty in fabrication. Even if we can acquire some of them, the work spectrums are mostly in a very narrow band [4]. By scarifying the material parameters (therefore some of the functionalities) into non-singular cases, broadband devices are proposed and realized (e.g., the carpet cloak [5,6,7,8,9,10]). Therefore, we see that these two factors, fancy effect and broadband functionality, are competing against each other, bringing about quite a dilemma for the field to advance.

In this article, we will prove that Fabry-Pérot (FP) resonances can address this difficulty to some extent. Firstly, we will propose a simple one dimensional singular mapping, which can be used to design a device called "optical void". We then set equivalency between such an optical void with the one dimensional metallic slit array. The working frequencies are right at the FP resonances. With this, we can design various devices, such as a concentrator, a shifter, a rotator, a waveguide bend, and a waveguide periscope. In particular, we fabricate a prototype of a concentrator for microwave frequencies.

Let us start with the simple mapping (see in Figure 1c):

$$x' = \begin{cases} x - x_1 + x_1' & x < x_1 \\ x_1' + \dfrac{x_2' - x_1'}{x_2 - x_1}(x - x_1) & x_1 \leq x < x_2 \\ x - x_2 + x_2' & x \geq x_2 \end{cases}$$
$$y' = y \tag{1}$$
$$z' = z$$

This mapping transformed a slab $x_1 \leq x < x_2$ in virtual space (Figure 1a) into a slab $x_1' \leq x < x_2'$ in physical space (Figure 1b). Suppose the virtual space is free space, we can obtain the material tensors of the slab $x_1' \leq x < x_2'$,

$$\vec{\varepsilon} = \vec{\mu} = \begin{pmatrix} \Delta & & \\ & \dfrac{1}{\Delta} & \\ & & \dfrac{1}{\Delta} \end{pmatrix} \tag{2}$$

where $\Delta = \dfrac{x_2' - x_1'}{x_2 - x_1}$. If $x_2 \to x_1$, $\Delta \to \infty$, the slab $x_1' \leq x < x_2'$ functions as an optical void, which means that when wave impinges, the phase does not accumulate after passing through it.

As an example, we plot the field pattern for a point source setting nearby the optical void in Figure 2a and choose the transverse magnetic (TM) polarization in this article, as we will later see that such a polarization with FP resonances has an advantage over the transverse electric (TE) polarization. Here only $\varepsilon_x = \Delta$ and $\varepsilon_y = \mu_z = 1/\Delta$ are required. In the simulation, we set $\Delta = 100000$, $x_2' - x_1' = 2$ and the wavelength $\lambda = 1$. The point source is a numerical test by setting the magnetic field $H_z$ as a constant at a tiny circular boundary. From the field pattern, we find that the image of the point source is perfectly transmitted from left to right, indicating the functionality of the optical void.

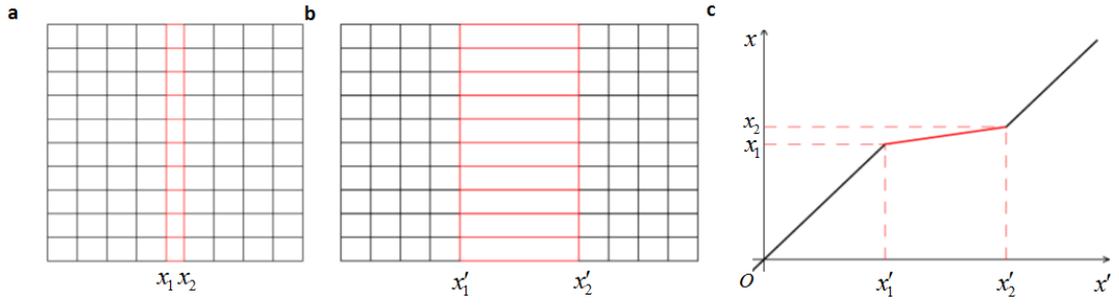

*Figure 1. The mapping of an optical void.* (a), The virtual space; (b) The physical space; (c) The detailed mathematical mapping.

Now we recall the one dimensional metallic slit array, which can support spoof surface plasmons [11]. Suppose the slit array is along $x$-direction, the width of the slits is $a$, the size of the unit cell is $d$. Through the design of the perfect endoscopes [12, 13, 14, 15, 16, 17], we know that when $x_2'-x_1'=m\frac{\lambda}{2}$ ($m=1,2,3...$), the same functionality of the optical void can be achieved. For example, as we set $d=2a=0.1$ in Figure 2b, the slit array can perfectly transmit an image of a point source at $\lambda=1$ (here $m=4$). As $\lambda \gg d$, the slit array can also be regarded as a slab with $\varepsilon_x=\infty$, $\varepsilon_y=2$ and $\mu_z=0.5$ [11]. The same functionality of this effective medium is demonstrated in Figure 2c (in the simulation $\varepsilon_x=100000$). In fact, we can get a general condition for such a perfect transmission, i.e.,

$$\int_{x_1'}^{x_2'}\sqrt{\varepsilon_y(x')\mu_z(x')}dx' = \int_{x_1'}^{x_2'} n(x')dx' = m\frac{\lambda}{2} \ (m=1,2,3...), \qquad (3)$$

which is actually the FP resonance condition.

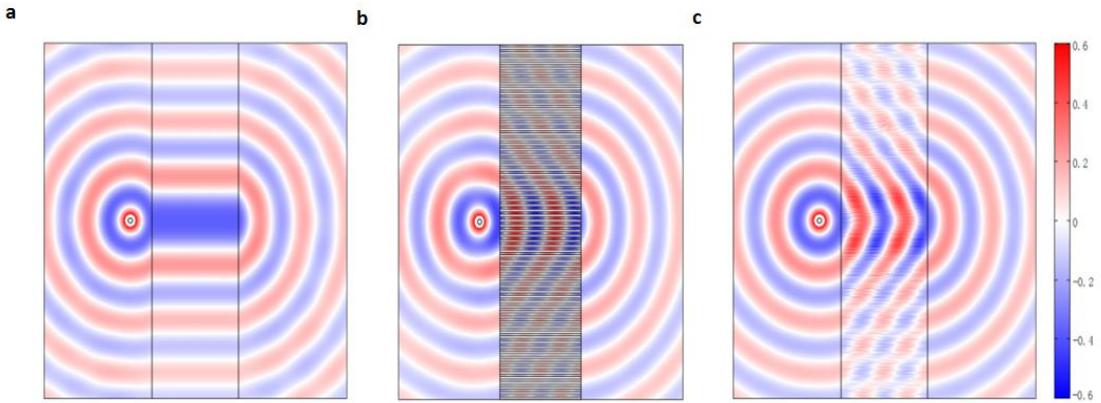

*Figure 2. The field pattern* for (a) an optical void; (b) the one dimensional metallic slit array; (c) effective medium for the slit array.

After the above equivalency is built up between the optical void and the metallic slit array, we can use the FP resonance in transformation optics and design many devices that can be easily implemented and capable of working for a series of frequencies.

For example, let us look at a similar mapping in circular cylindrical coordinate (see in Figure 3c) [18],

$$r' = \begin{cases} \dfrac{r_1'}{r_1} r & 0 \leq r < r_1 \\ r_1' + \dfrac{r_2' - r_1'}{r_2 - r_1}(r - r_1) & r_1 \leq r < r_2 \\ r & r \geq r_2 \end{cases}$$

$$\theta' = \theta \qquad\qquad (4)$$
$$z' = z$$

which maps a concentric cylindrical layer $r_1 \leq r < r_2$ in virtual space (Figure 3a) into another concentric cylindrical layer $r_1' \leq r' < r_2'$ in physical space (Figure 3b). This is a mapping to design a concentrator if we set virtual space as free space, $r_2' = r_2$ and $r_1' < r_1$. If $r_1 \to r_2$, we shall have $\varepsilon_r \to \infty$, $\varepsilon_\theta \to 0$ and $\mu_z \to 0$ for $r_1' \leq r' < r_2'$, and $\varepsilon = 1$ and $\mu_z = (\dfrac{r_1}{r_1'})^2$ for $0 \leq r' < r_1'$. Figure 4a shows the perfect transparency of such a concentrator for $\lambda = 1$, where we set $r_2' = r_2 = 2$, $r_1' = 1$ and $r_1 = 1.9999$.

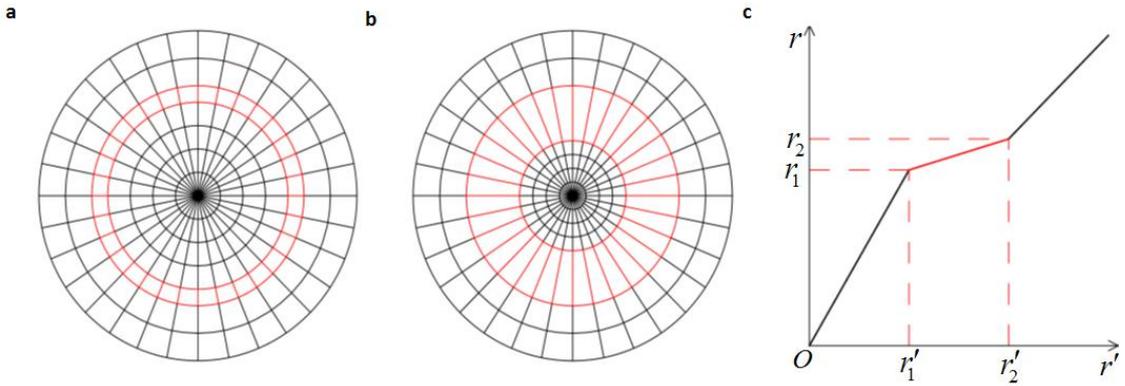

*Figure 3. The mapping of a cylindrical concentrator. (a), The virtual space; (b) The physical space; (c) The detailed mathematical mapping.*

Inspired by the above FP resonance in $x$-direction, here we can also design a concentrator from FP resonance in $r$-direction. For example, we can set $\varepsilon_r = \infty$, $\varepsilon_\theta = \varepsilon_\theta(r')$ and $\mu_z = 1$ for

$r_1' \leq r' < r_2'$, and $\varepsilon = (\frac{r_2'}{r_1'})^2$ and $\mu_z = 1$ for $0 \leq r' < r_1'$. The FP resonance condition in $r$-direction is:

$$\int_{r_1'}^{r_2'} \sqrt{\varepsilon_\theta(r')\mu_z}\, dr' = \int_{r_1'}^{r_2'} n(r')dr' = m\frac{\lambda}{2} \quad (m=1,2,3...), \quad (5)$$

Such a concentrator does not have any magnetic response and can work for a series of frequencies. To show the similar perfect transparency, we plot the field pattern in Figure 4b for $\varepsilon_\theta(r') = 3 - r'$ and $\lambda = 1$ (here $m = 3$, $r_2' = 2$, $r_1' = 1$). We note that such a version is a perfect one as the impedances match at both boundaries ($r' = r_1'$ and $r' = r_2'$). How can we realize such a concentrator? We can simply insert thin metallic plates along $r$-direction in a dielectric profile $\varepsilon(r')$ in $r_1' \leq r' < r_2'$. For example, we insert *144* pieces of metallic plates and plot the field pattern for $\lambda = 1$ in Figure 4c, where almost perfect transparency can be observed. The effect medium in $r_1' \leq r' < r_2'$ should be modified to $\varepsilon_r = \infty$, $\varepsilon_\theta = \varepsilon(r')/(1-f)$ and $\mu_z = 1-f$. In this simulation, $f = 0.2$. If $f$ is too large, impedances will mismatch at both boundaries ($r' = r_1'$ and $r' = r_2'$), causing some scattering and imperfection.

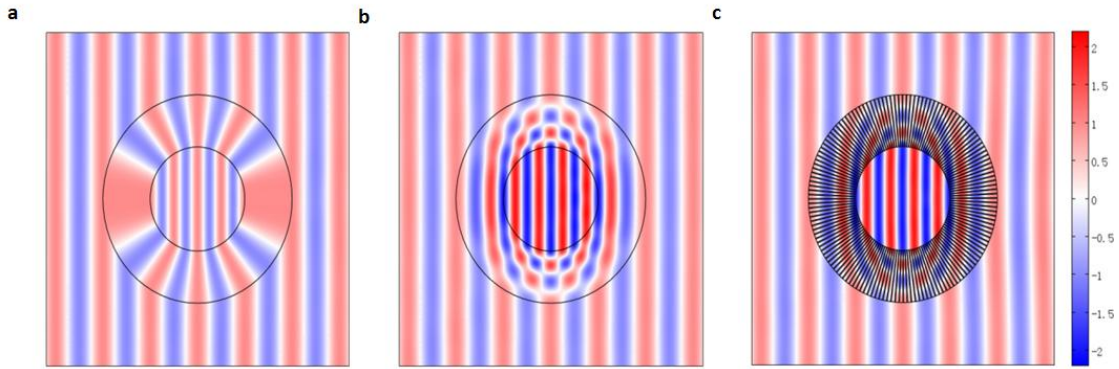

*Figure 4. The field pattern for cylindrical concentrators.* (a), The original version; (b), The FP version; (c) Implementations by inserting thin metallic plates in a dielectric profile.

Now we come to implement a reduced version of the above Figure 4c. As it is still quite tedious to realize the part between $r_1' \leq r' < r_2'$. We make further reduction, that is, to just insert the metallic plates in air. We find from simulation that, apart from a little scattering induced by the impedance mismatching at the inner boundary $r' = r_1'$, the concentrating effect is still there. To get a perfect transparency like that in Figure 4c, we can simply replace the dielectric core ($\varepsilon = (\frac{r_2'}{r_1'})^2$ and $\mu_z = 1$) with a metamaterial core ($\varepsilon = \mu_z = \frac{r_2'}{r_1'}$).

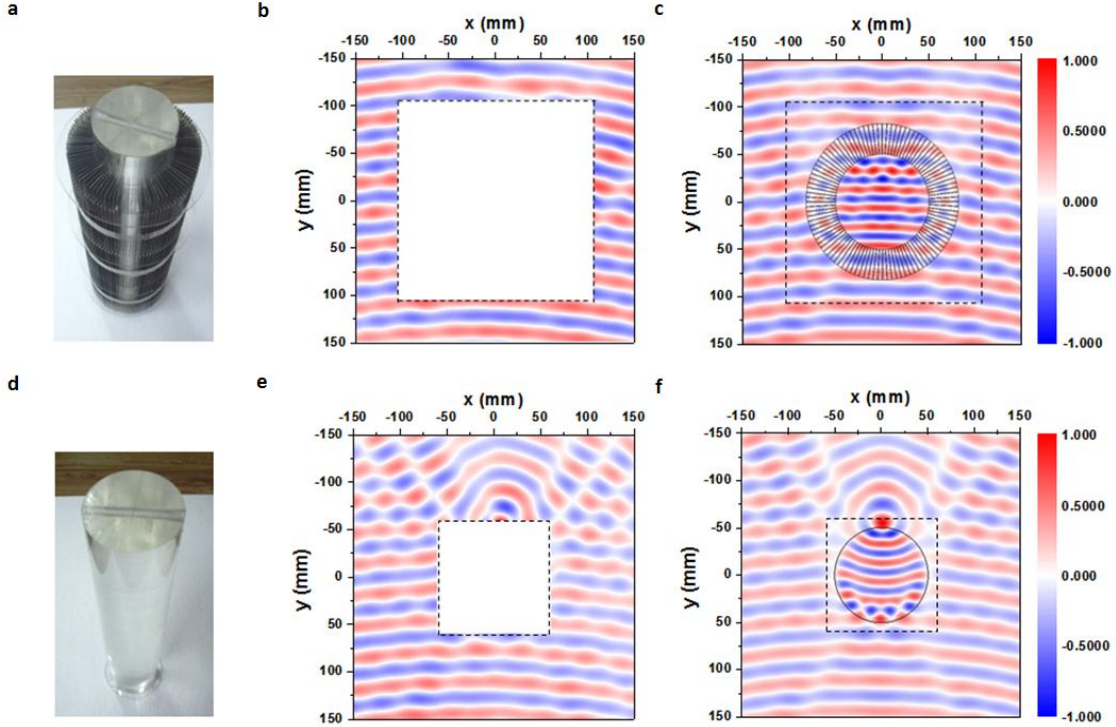

*Figure 5. The concentrator and the effect.* (a), The photo of the concentrator with the solid Plexiglas cylinder; (b), The measured $H_z$ field immediately neighboring the concentrator at 9.26GHz; (c), The simulated result; (d), The photo of the bare Plexiglas cylinder; (e), The measured $H_z$ field immediately neighboring the cylinder at 9.26 GHz; (f), The simulated result.

The concentrator is assembled using the thin iron pieces, with an annular cross section of $50 \text{ mm} < r' < 82 \text{ mm}$ in its final form, and is fixed surrounding a solid Plexiglas cylinder with a dielectric constant *2.7* (see in Figure 5a). Figure 5b is the measured $H_z$ field at *9.26GHz*. The scanning range measures *300×300 mm²*, starting with the upper left corner, and the quasi-plane wave with slightly curved wavefronts is incident towards the *-y*-direction. The central area in the scanning region is not accessible due to the sample occupation. It is seen that the wave recovers its original wavefronts with negligible intensity decay after passing through the sample, which points out the scattering free feature of our device. To show the full picture of the $H_z$ field, we show the simulation result in Figure 5c, where the concentrating effect is well demonstrated with a little scattering. For comparison, we also present the results for the bare Plexiglas cylinder (photo taken in Figure 5d). The measured $H_z$ field at *9.26GHz* is shown in Figure 5e, and the related simulated result is shown in Figure 5f, where we find that the scattering is enlarged. From the simulation, the dielectric core also shows a focusing effect. But with the concentrator, the wavefronts inside it become quite flat, as shown in Figure 5c.

In fact, from the simulation, we find that the best working frequency is at *9.5GHz*. But due to the imperfection of the sample fabrication, the working frequency is shifted to *9.26GHz*. In figure 6, we plot the frequency dependences of the total scattering cross sections for the bare dielectric core and the concentrator with metallic plates. We find that the concentrator here works well for multiple frequencies. Near the resonant frequencies, the scattering cross sections are smaller than those of the bare dielectric core.

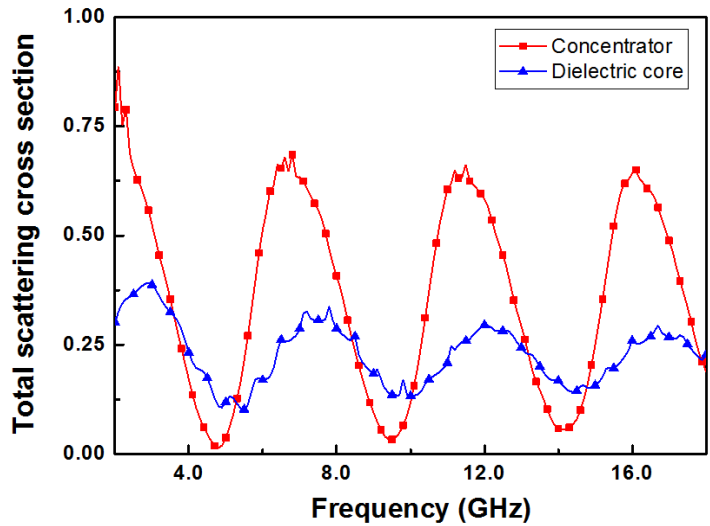

*Figure 6. The total scattering cross sections* of the bare dielectric core and the concentrator with metallic plates at different frequencies.

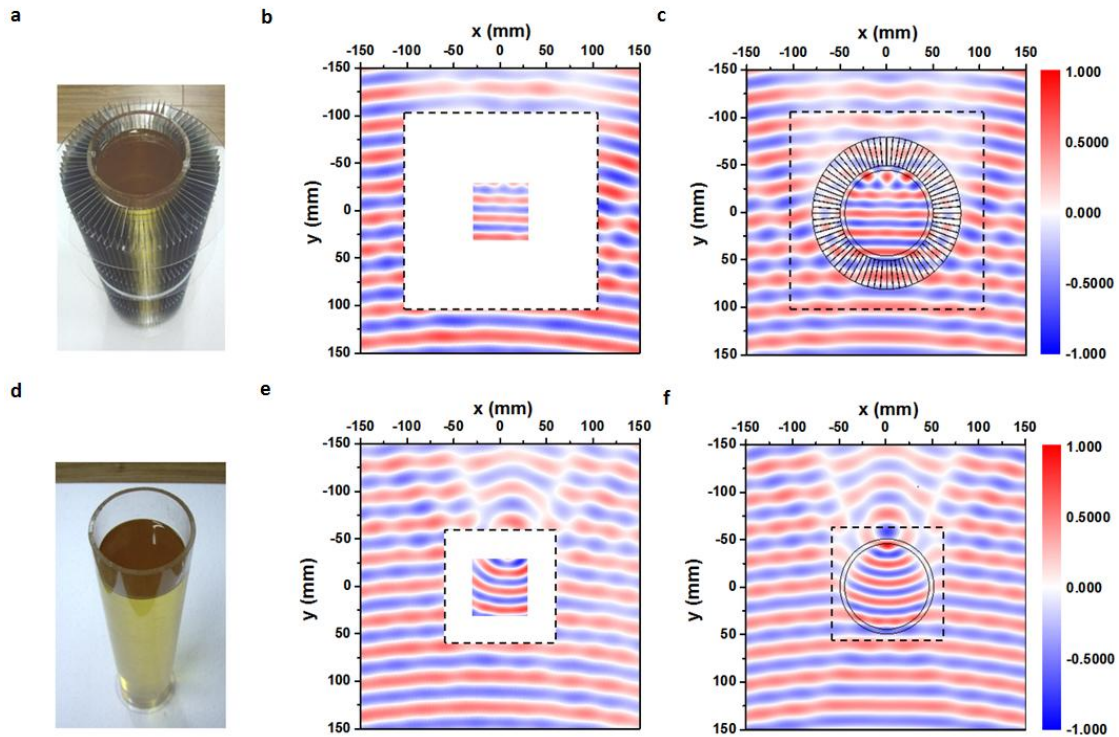

*Figure 7. The concentrator with oil cylinder and the effect.* (a), The photo of the concentrator

*with the oil cylinder; (b), The measured $H_z$ field immediately neighboring the concentrator and inside the oil at 9.35GHz; (c), The simulated result; (d), The photo of the bare oil cylinder; (e), The measured $H_z$ field immediately neighboring the cylinder and inside the oil at 9.35 GHz ; (f), The simulated result.*

In order to detect the field inside the core region in the concentrator, we employed a liquid sample, as illustrated in Figure 7a. A cylindrical Plexiglas container with the inner radius *45mm*, the outer radius *50mm*, and the height *53cm* is filled with castor oil, as shown in Figure 7d. Because the castor oil has the dielectric constant *2.6+0.1i*, close to the Plexiglas's value *2.7*, the oil loaded container is considered to be a liquid cylinder. The associated concentrator is fabricated using *72* iron pieces, and has an annular cross section of $50 \text{ mm} < r' < 80 \text{ mm}$ in its final form. Besides scanning the exterior field, we insert the detector into the oil to sense the interior field inside the oil cylinder. A central square in the cylindrical container is now accessible to our detector. Figure 7b and 7c are the measured and simulated $H_z$ field at *9.35GHz*, respectively. The field patterns show that the scattering is small and the wavefronts in the oil are flat, which turns out to coincide with the concentrating effect as predicted. For comparison, we also present the results for the bare oil cylinder in Figure 7d. The measured $H_z$ field and the simulated result at *9.35GHz* are shown in Figure 7e and 7f, respectively. From the measured field pattern in the oil cylinder, the focusing effect is seen clearly, and is consistent with the simulated result. Due to the imperfection of the sample fabrication and the dissipation in the oil core, the working frequency is now shifted from *10GHz* to *9.35GHz*.

Apart from the above implementation of a reduced FP concentrator, we also design a shifter [19], a rotator [20], a waveguide bend [21], and a waveguide periscope in the Supplementary Figures. Given the same principle as the above, we only numerically show the refractive index profiles and the related functionalities. Supplementary Figure S1(a) shows the index profile of the shifter, which is composed of oblique metallic slit arrays. Supplementary Figure S1(b) shows that the beam incident to the shifter has a displacement in $y$-direction after passing through it. By inserting specially designed curved metallic structures into an index profile in Supplementary Figure S2(a), a perfect rotator can be implemented and the functionality is shown in Supplementary Figure S2(b). Thirdly, we can implement devices by combining the optical conformal mapping [1] together with the FP resonances. As the optical conformal mapping (from $z = x + iy$ to $w = u + iv$) keeps the optical paths unchanged, i.e., $\int n_w |dw| = \int n_z |dz|$, if we insert curved metallic structures along $u$-lines (or $v$-lines), the phase changes of any waves at $v$-lines (or $u$-lines) will be unchanged, therefore all the paths have the same FP resonant frequencies. Take $w = 2\ln z$ as an example, we can design a waveguide bend. The index profile is shown in Supplementary Figure S3(a) (with the metallic structures along the $\theta$-direction), while the related functionality is proved in Supplementary Figure S3(b). Likewise, if we use the Zhukowski mapping, i.e., $w = z + 1/z$, a perfect one dimensional cloak can be achieved (not shown in this article). Finally, we show a perfect waveguide periscope in Supplementary Figure S4 (a) by simply inserting a specially designed curved metallic structure in air. The functionality is demonstrated in Supplementary Figure S4 (b). In fact, the concentrator here can also be used as a hyperlens [22], or even find some potential application in energy harvests or wireless power transfer.

In conclusion, we have shown that FP resonances can help to design various transformation optical devices, which can not only be easily implemented as they are only composed of (curved if necessary) metallic structures and dielectric profile, but can also work for a series of frequencies. As an example, we fabricate a prototype for a concentrator in microwaves and demonstrate its functionalities.

**Method**

**Theory and simulation.** All the simulated field patterns and the scattering cross sections in Figure 6 are obtained using the finite element solver COMSOL Multiphysics.

**Sample fabrication.** The field concentrator is a cylindrical object comprising thin iron sheets arranged into the theoretically designed pattern. The iron sheet has the thickness *0.3 mm*, and is cut into the $32 \times 500$ *mm*$^2$ rectangular piece. We fabricate a total of *100* pieces and assemble them to be the concentrator. Their relative positions were fixed via inserting the iron pieces into an annular Plexiglas slice with designed radial slits. With several Plexiglas slice fixtures, we fabricate the field concentrator which has an annular cross section of $50 \text{ mm} < r' < 82 \text{ mm}$ and a height of *500 mm*. To facilitate the measurement, the second concentrator with *72* iron pieces and the size $50 \text{ mm} < r' < 80 \text{ mm}$ is made to be used together with the oil cylinder.

**Experimental set-up.** In the experiment, the samples are placed vertically with the cylindrical axis orientated along the *z*-direction, and a horn-shaped antenna, located *~100 cm* away from the field concentrator, transmits microwaves toward the sample with the H-field polarized along the vertical direction, as depicted in Supplementary Figure S5. To map the spatial distribution of the $H_z$ component, we employ a split ring detecting antenna which is made of a coaxial cable and has a circular loop of diameter *4 mm* and a split *1 mm*. Its *S11* spectrum is measured and shown in Supplementary Figure S6, displaying a magnetic radiating/receiving characteristic around *10GHz*. To detect the magnetic field inside the oil sample, the ring plane of the detector is perpendicular to the *z*-directed coaxial cable which is inserted into the oil from the top of the sample. The split-ring antenna as well as the coaxial cable is mounted on a two-dimensional translation stage and is controlled to move in the horizontal *x-y* plane. The scanning range covers a square of *300×300 mm*$^2$ with a spatial resolution of *2×2 mm*$^2$, but the central area is not accessible due to the sample occupation, as marked by the dashed lines in Supplementary Figure S5. The horn antenna and the detector are connected to an S-parameter network analyzer to obtain the magnitude and phase of the $H_z$ field.

**Measurements.** Before the measuring the field patterns of the scatters or concentrators, we scan the magnitude and phase of the incident $H_z$ field in air at *9.26GHz* in Supplementary Figure S7(a) and (b). Supplementary Figure S7(c) shows the real part of $H_z$ field, which has the quasi-plane wave feature with slightly curved wavefronts. The related simulated magnitude, phase, and the real part of the incident $H_z$ field are shown in Supplementary Figure S7(d), (e) and (f). Likewise, Supplementary Figure S8(a), (b), and (c) show the magnitude, phase, and the real part of incident $H_z$ field in air at *9.35GHz,* while the related simulated magnitude, phase, and the real part are shown in Supplementary Figure S8(d), (e) and (f). To get the field pattern of the bare solid Plexiglas cylinder in Figure 5e (or Supplementary Figure S9(c)), we scan the magnitude and phase of the $H_z$ field in Supplementary Figure S9(a) and (b). The simulated magnitude, phase, and the real part of the $H_z$ field are shown in Supplementary Figure S9(d), (e) and (f). Likewise, we measure the magnitude and phase of the $H_z$ field for the concentrator with the solid Plexiglas cylinder in Supplementary Figure S10(a) and (b), which can be used to obtain the real part of the $H_z$ field in Supplementary Figure S10(c) (or Figure 5c). The simulated magnitude, phase, and the real part of the $H_z$ field for the concentrator are shown in Supplementary Figure S10(d), (e) and (f). For the bare oil cylinder, the magnitude, phase, and the real part of the $H_z$ field are shown in Supplementary Figure S11(a), (b) and (c) (or Figure 7e),while the simulated magnitude, phase, and the real part are shown in Supplementary Figure S11(d), (e) and (f). For the concentrator with the oil cylinder, the magnitude, phase, and the real part of the $H_z$ field are shown in Supplementary Figure S12(a), (b) and (c) (or Figure 7c), while the simulated magnitude, phase, and the real part are shown in Supplementary Figure S12(d), (e) and (f).

**Acknowledgements**

This work is supported by National Science Foundation of China for Excellent Young Scientists (grant no. 61322504), the Foundation for the Author of National Excellent Doctoral Dissertation of China (grant no. 201217), the National Natural Science Foundation of China (grant nos. 11004147 and 11104198), and the Priority Academic Program Development (PAPD) of Jiangsu Higher Education Institutions. H. C. would like to thank Winsley Yang's effort to polish the English.

**Author contributions**

H. C. conceived the idea. H. C., M. M. S., and L. X. did the theoretical calculations and the numerical simulations. B. H., S. L., and L. X. fabricated the samples and did the experimental measurements. B. H. supervised the experimental part. H. C. supervised the whole project. H .C. and B. H. wrote the manuscript.

**Additional information**

The authors declare no competing financial interests. Correspondence and requests for materials should be addressed to H. C., or to B. H.

**Supplementary Figures**

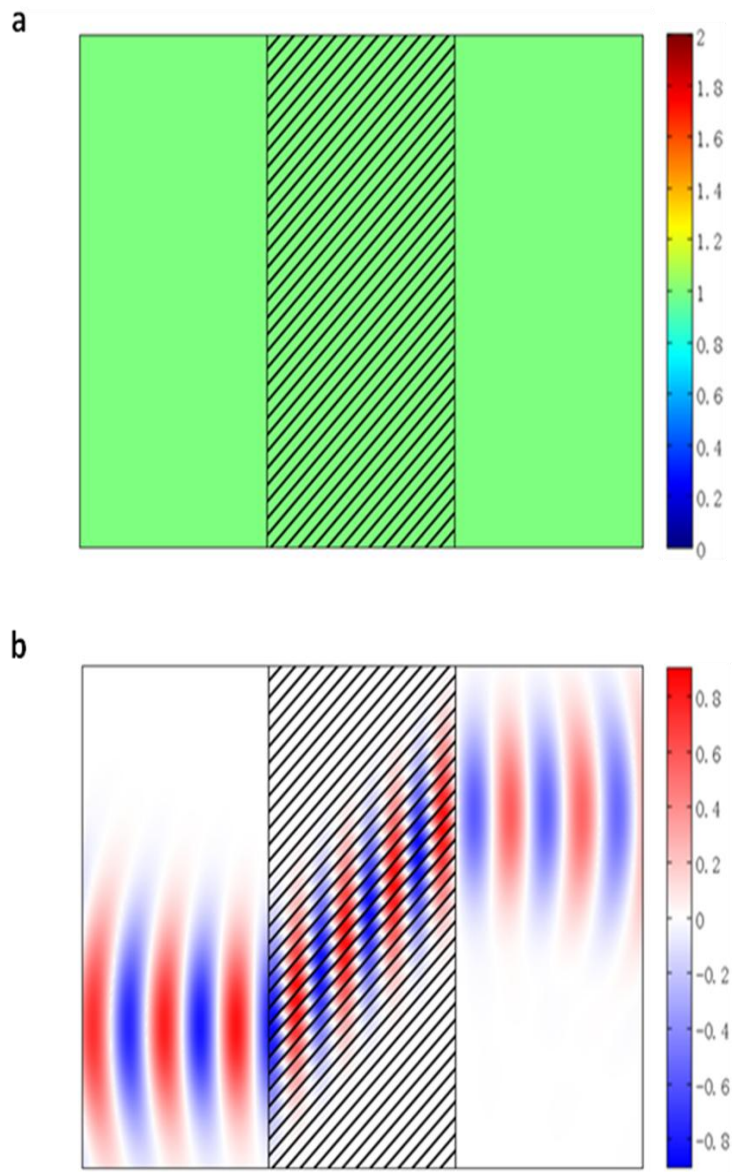

**Supplementary Figure S1. The shifter with FP resonance.** (a), The refractive index profile; (b), The shifting functionality.

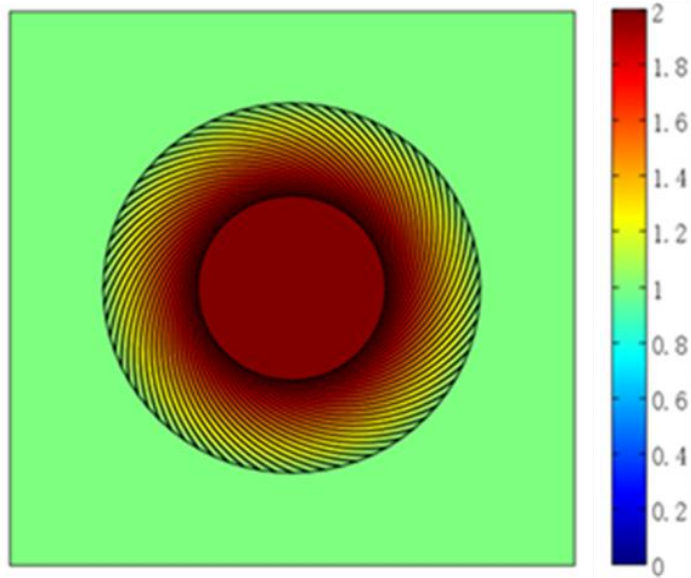

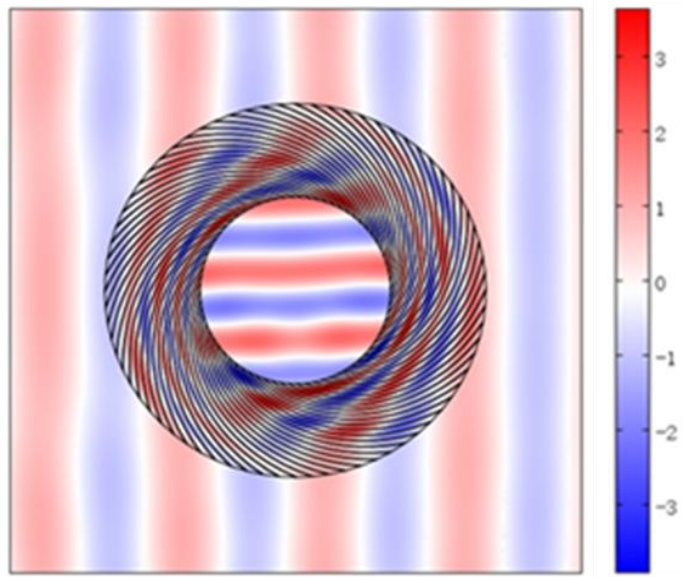

**Supplementary Figure S2. The rotator with FP resonance.** (a), The refractive index profile; (b), The wavefront rotation functionality.

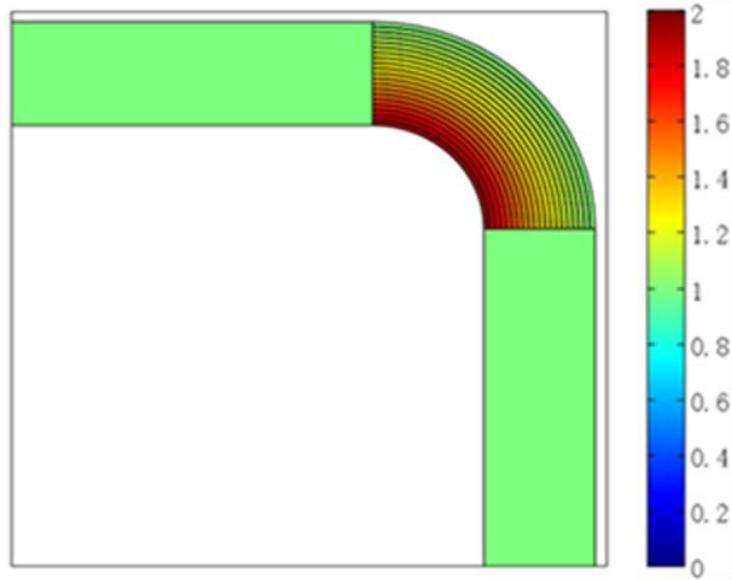

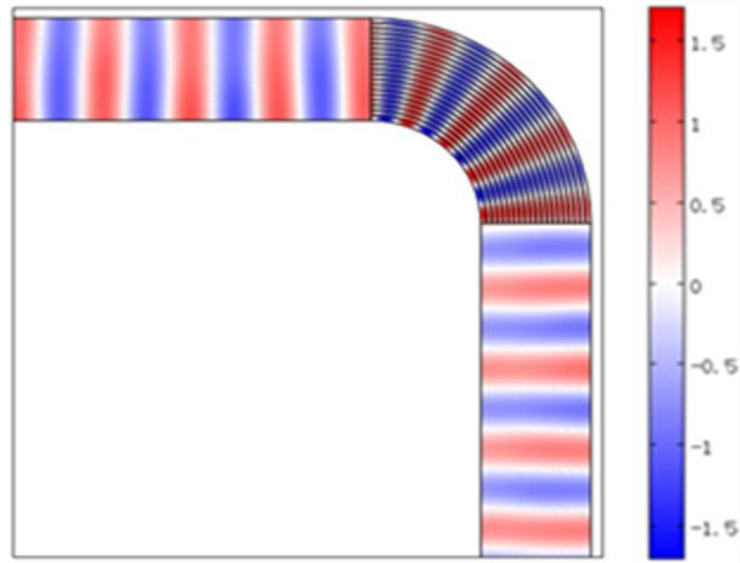

**Supplementary Figure S3. The waveguide bend with FP resonance.** (a), The refractive index profile; (b), The wave bending functionality.

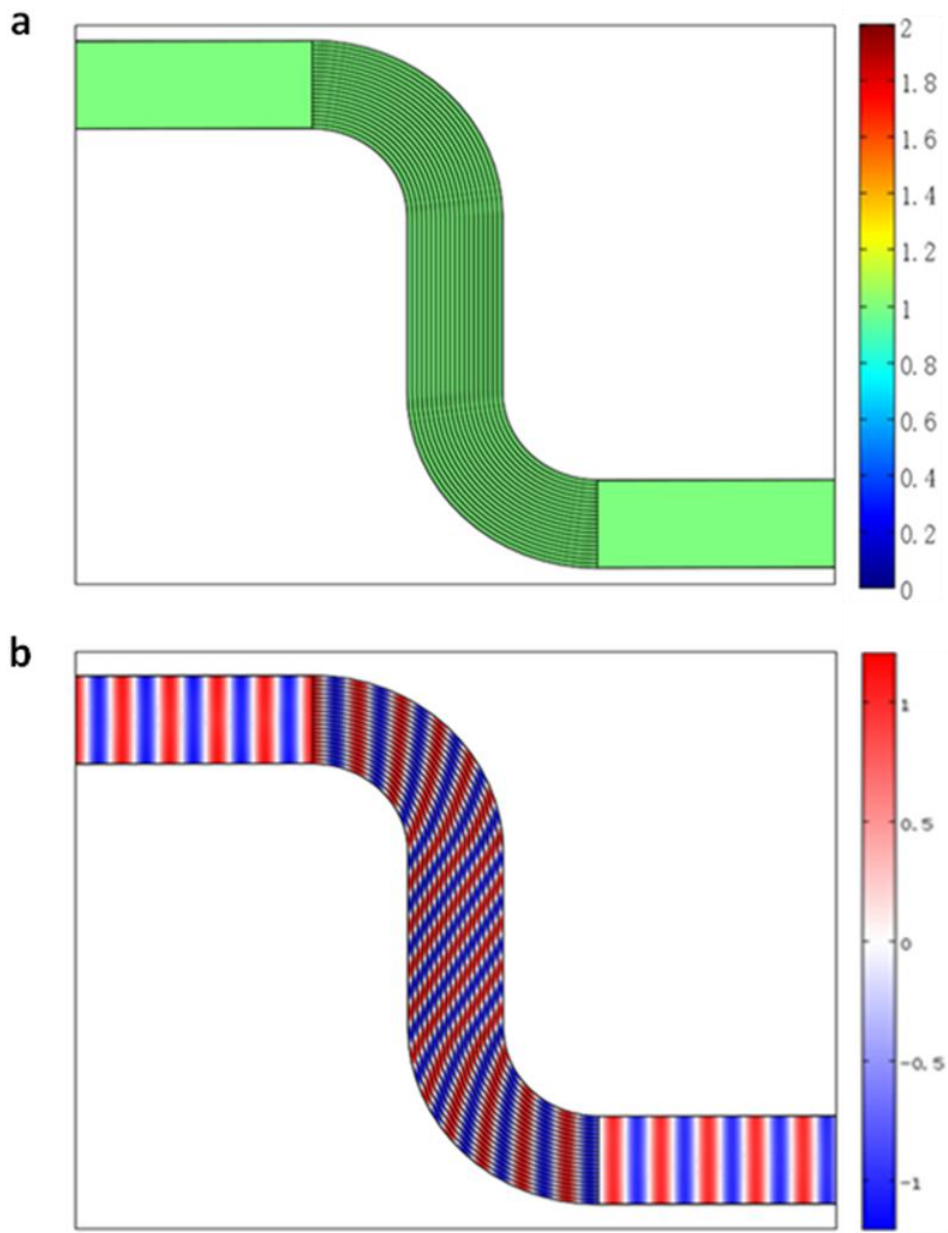

**Supplementary Figure S4. The waveguide periscope with FP resonance.** (a), The refractive index profile; (b), The periscope functionality.

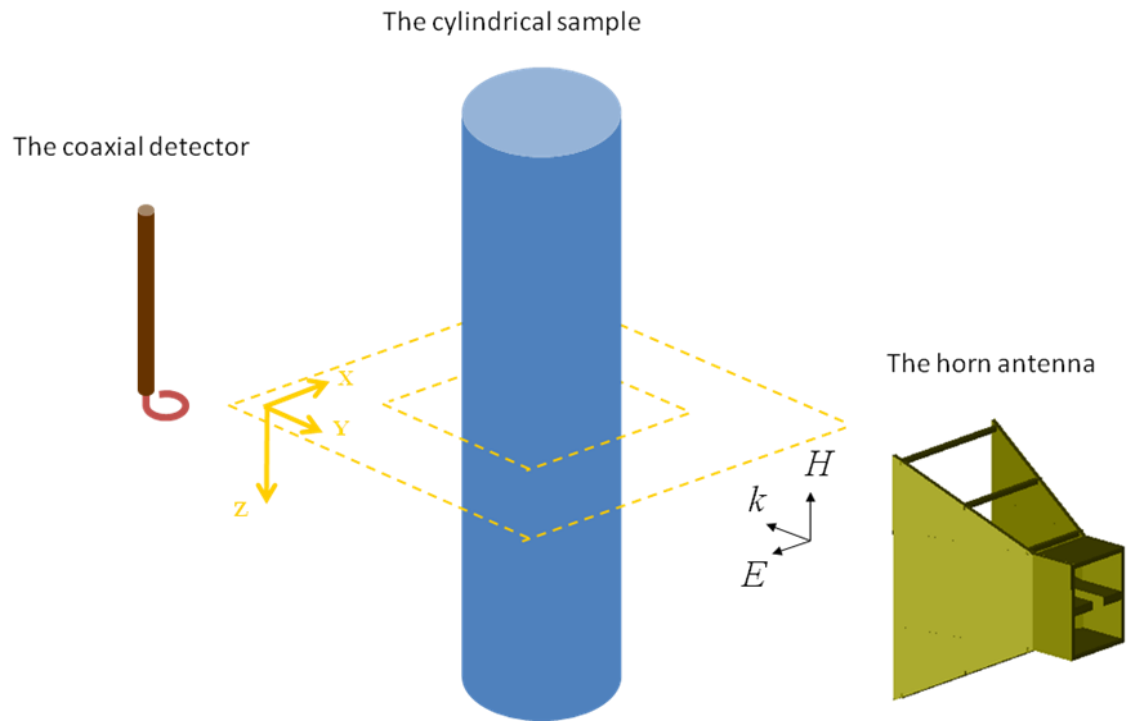

**Supplementary Figure S5. The measurement setup.** In the experiment, the samples are placed vertically with the cylindrical axis orientated along the *z*-direction, and a horn-shaped antenna, located ~*100 cm* away from the samples, transmits microwaves toward them with H-field polarized along the vertical direction. The scanning plane, measuring *300×300 mm$^2$*, is located in the middle of the *500 mm* high samples to avoid the ending effect.

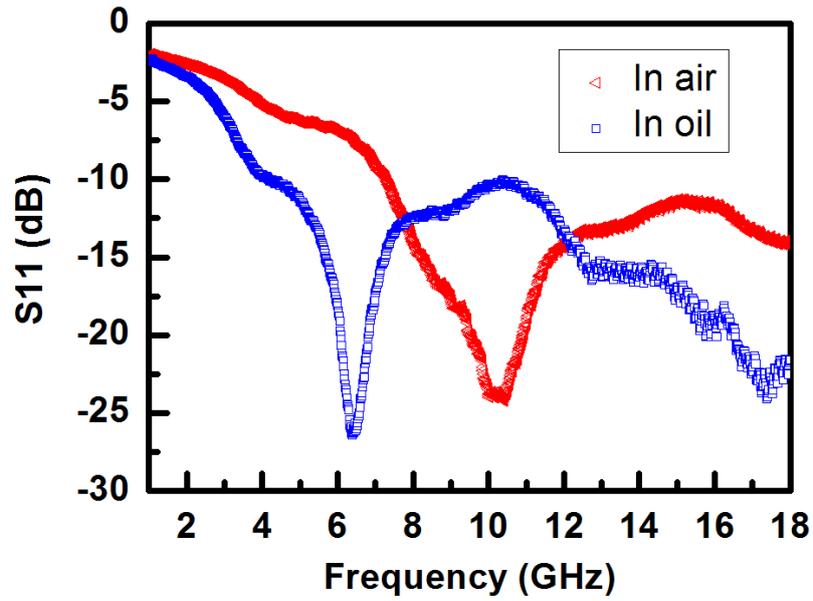

**Supplementary Figure S6. The magnetic response of the detector.** To map the spatial distribution of the $H_z$ component, we employ a split ring detecting antenna which is made of a coaxial cable and has a circular loop of diameter *4 mm* and a split *1 mm*. Its *S11* parameter is characterized first to identify the magnetic response frequency which is shown around *10GHz* in air. When the detector is inserted into the oil, the overall spectrum will downshift, bringing about a change in the radiating/receiving capability of the detector at each specific frequency, for example, one of our working frequencies (*9.35GHz*). Thus, the scanned field magnitude inside the oil cylinder needs to be corrected to make the direct comparison with the external fields in air.

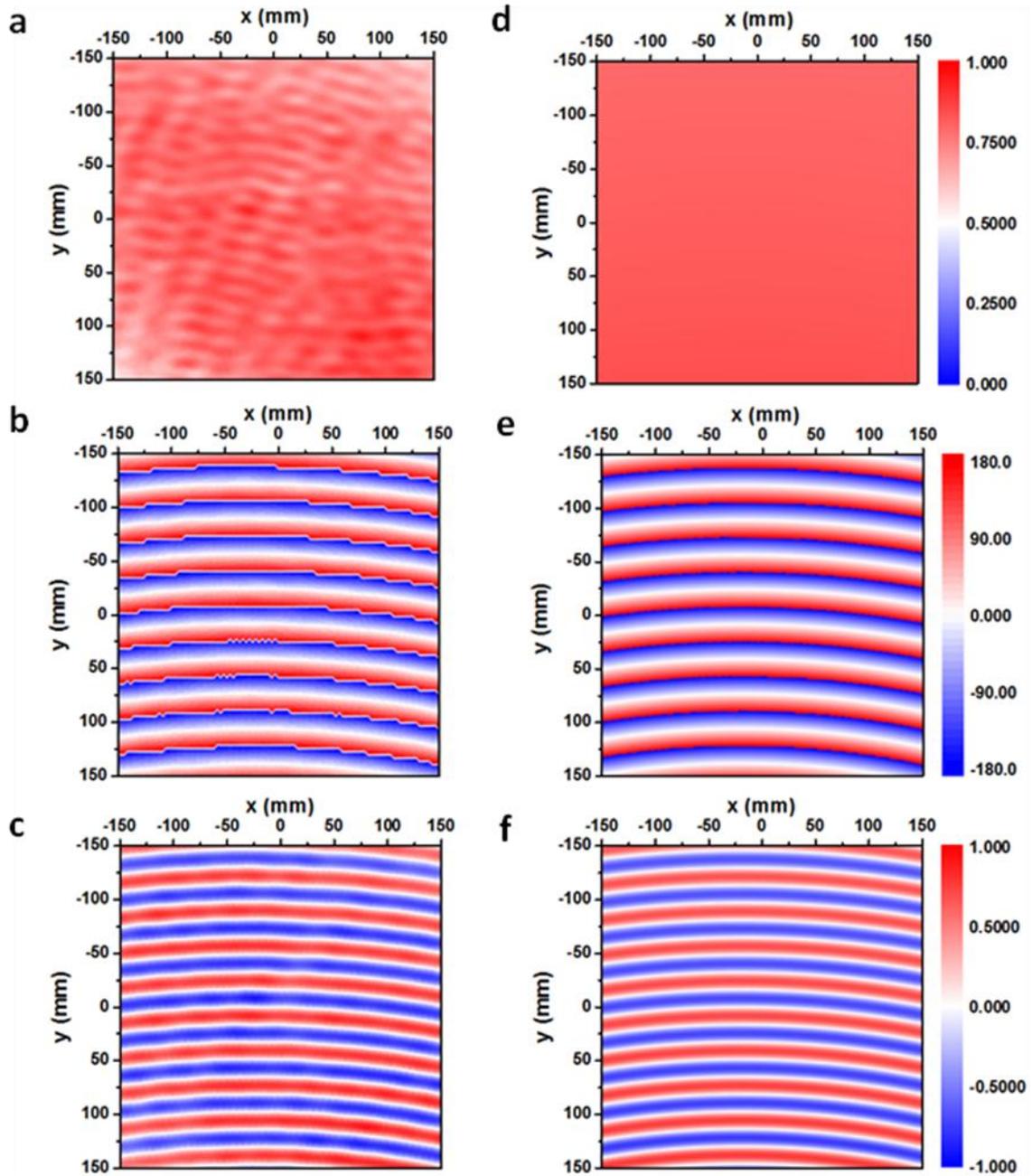

**Supplementary Figure S7. The incident H-field at *9.26GHz* in free space.** (a), The magnitude of the measured field; (b), The phase of the measured field; (c), The real part of the measured field; (d), The magnitude of the simulated field; (e), The phase of the simulated field; (f), The real part of the simulated field. It is seen that the field is incident toward the *-y* direction and has the quasi-plane wave feature with slightly curved wavefronts.

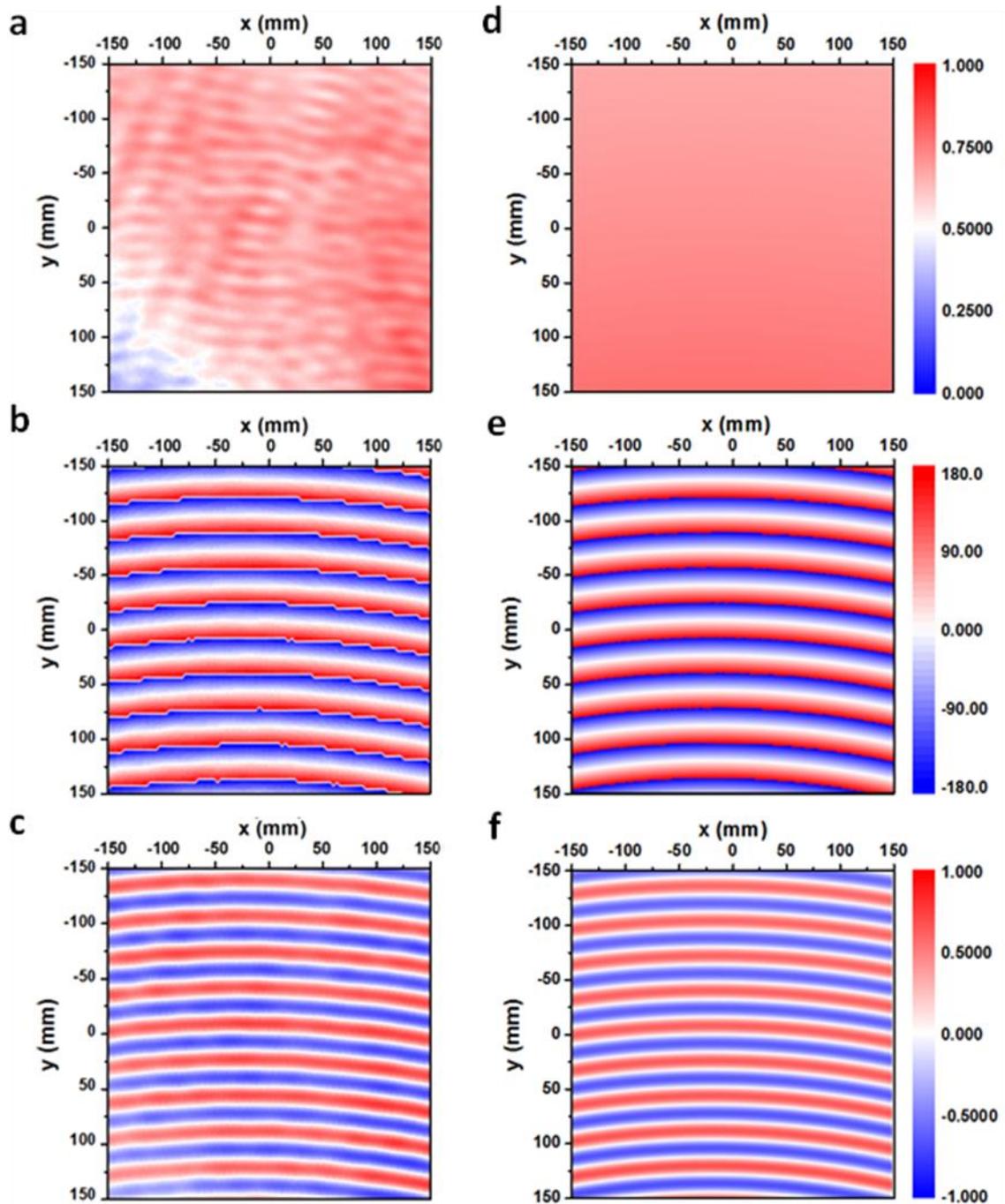

**Supplementary Figure S8. The incident H-field at *9.35GHz* in free space.** (a), The magnitude of the measured field; (b), The phase of the measured field; (c), The real part of the measured field; (d), The magnitude of the simulated field; (e), The phase of the simulated field; (f), The real part of the simulated field.

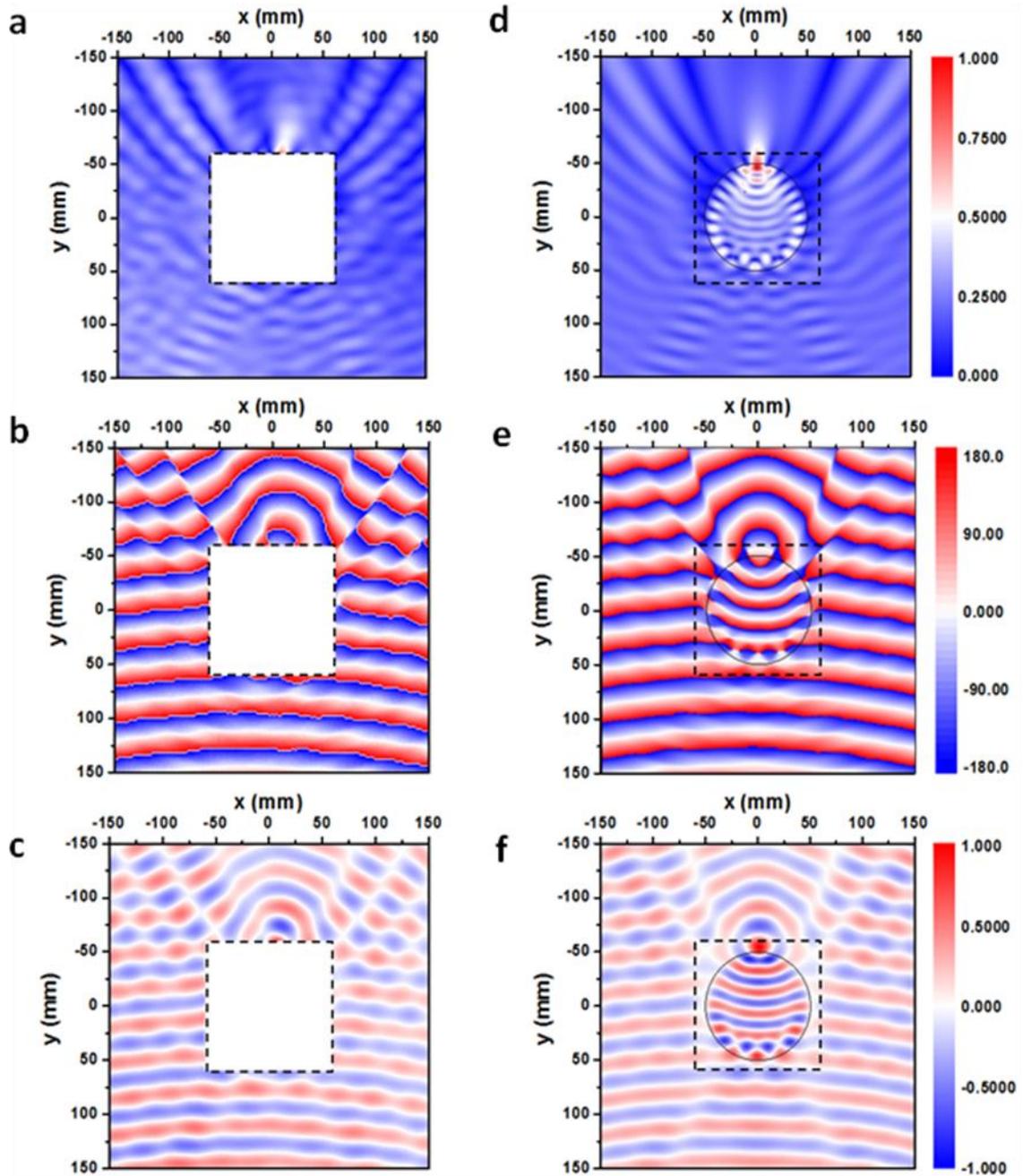

**Supplementary Figure S9. The H-field of the solid Plexiglas cylinder at *9.26GHz*.** (a), The magnitude of the measured field; (b), The phase of the measured field; (c), The real part of the measured field; (d), The magnitude of the simulated field; (e), The phase of the simulated field; (f), The real part of the simulated field.

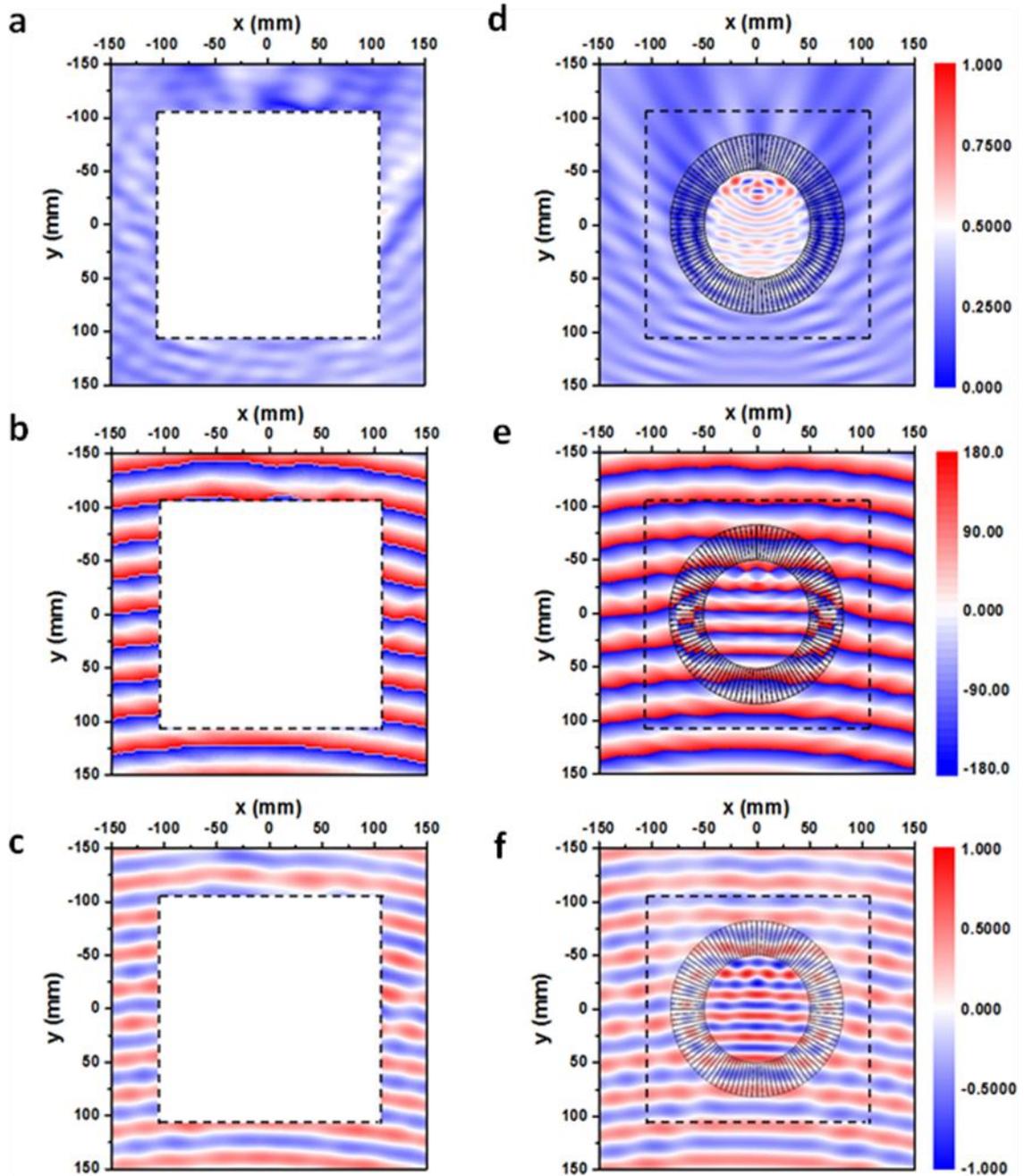

**Supplementary Figure S10. The H-field of the concentrator with the solid Plexiglas cylinder at *9.26GHz*.** (a), The magnitude of the measured field; (b), The phase of the measured field; (c), The real part of the measured field; (d), The magnitude of the simulated field; (e), The phase of the simulated field; (f), The real part of the simulated field.

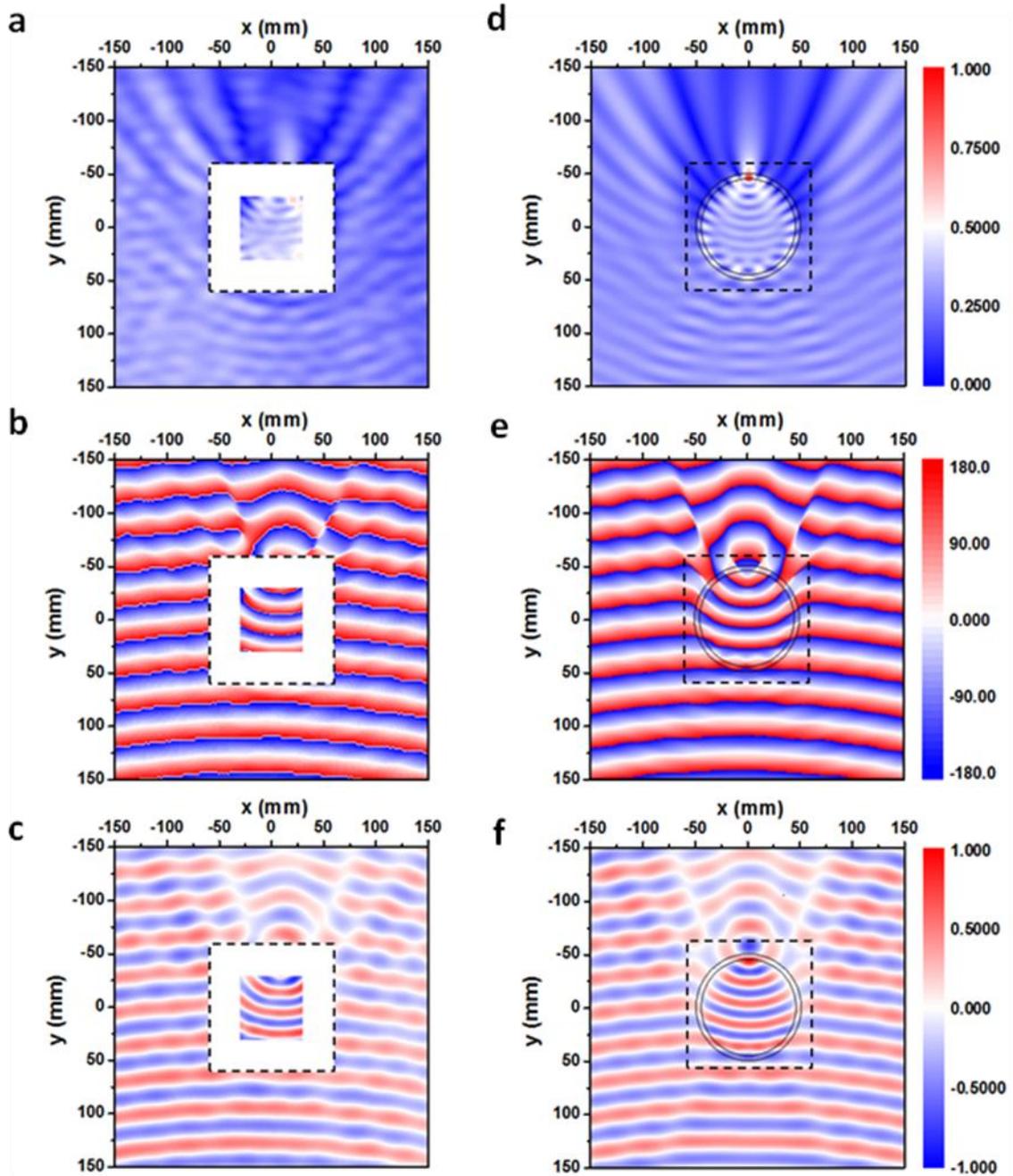

**Supplementary Figure S11. The H-field of the oil cylinder at *9.35GHz*.** (a), The magnitude of the measured field; (b), The phase of the measured field; (c), The real part of the measured field; (d), The magnitude of the simulated field; (e), The phase of the simulated field; (f), The real part of the simulated field.

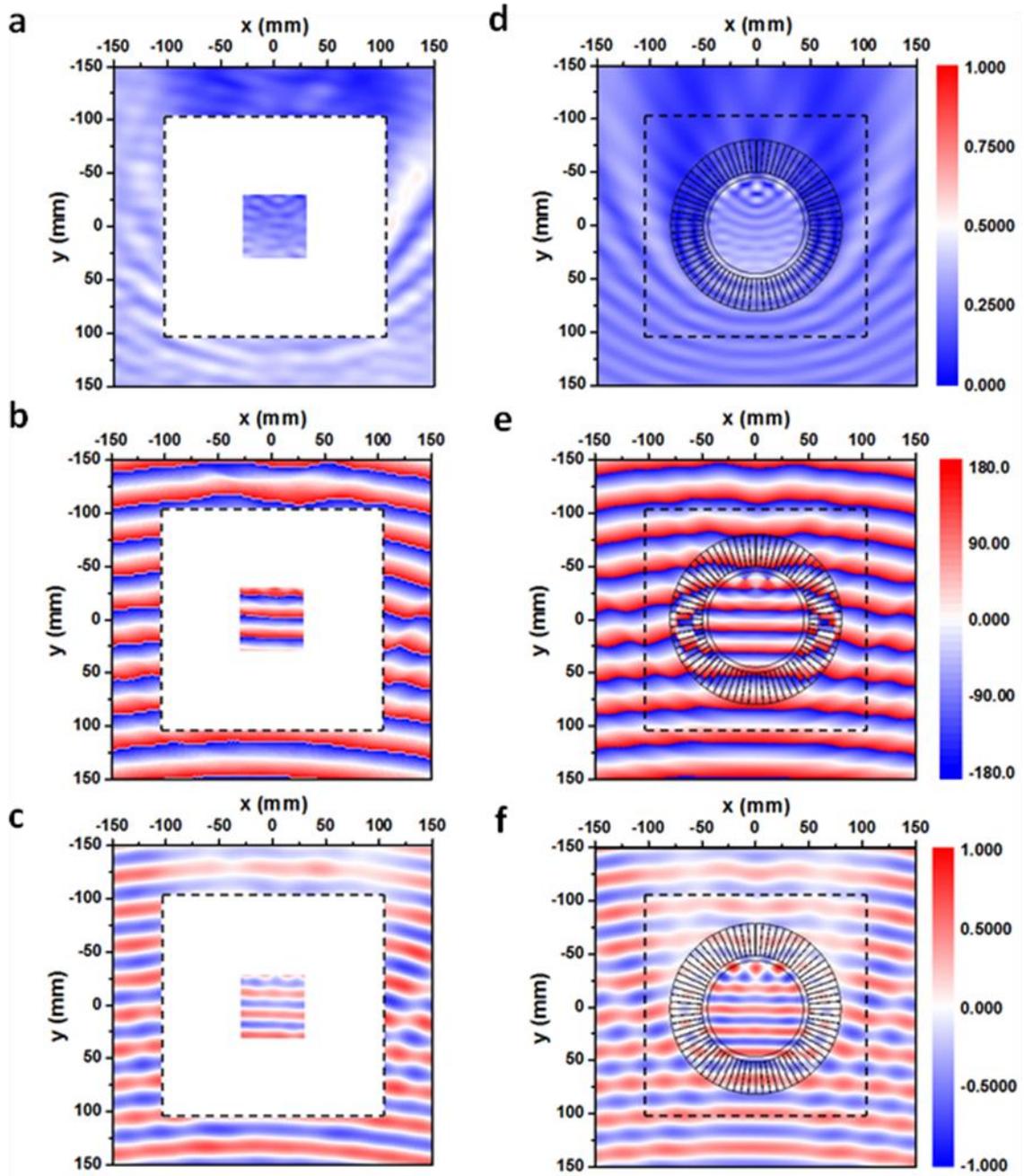

**Supplementary Figure S12. The H-field of the concentrator with the oil cylinder at *9.35GHz*.** (a), The magnitude of the measured field; (b), The phase of the measured field; (c), The real part of the measured field; (d), The magnitude of the simulated field; (e), The phase of the simulated field; (f), The real part of the simulated field.